\documentclass[aps,prl,twocolumn,showpacs]{revtex4}
\usepackage{amssymb}
\usepackage{amsmath}
\usepackage{amsthm}
\usepackage{graphicx}
\usepackage{dcolumn}
\usepackage{bm}

\begin{document}

\title{Quantum correlation measure in arbitrary bipartite systems}
\author{ Chang-shui Yu$^1$}
\email{quaninformation@sina.com; ycs@dlut.edu.cn}
\author{ Shao-xiong Wu$^1$}
\author{ Xiaoguang Wang$^2$}
\author{ X. X. Yi$^1$}
\author{He-shan Song$^1$}
\affiliation{$^1$School of Physics and Optoelectronic Technology, Dalian University of
Technology, Dalian 116024, China }
\affiliation{$^2$Zhejiang Institute of Modern Physics, Department of Physics, \\
Zhejiang University, Hangzhou 310027, China }
\date{\today }

\begin{abstract}
A definition of quantum correlation is presented for an arbitrary bipartite
quantum state based on the skew information. This definition not only
inherits the good properties of skew information such as the contractivity
and so on, but also is effective and almost analytically calculated for any
bipartite quantum states. We also reveal the relation between our measure
and quantum metrology. As applications, we give the exact expressions of
quantum correlation for many states, which provides a direct support for our
result.
\end{abstract}

\maketitle

\textit{Introduction.}-When quantum correlation is mentioned, one might be
immediately tempted to think of quantum entanglement which results from the
merging of the superposition principle of states and the tensor structure of
the composite quantum state space. This is not strange because as one of the
most intriguing feature of quantum mechanics that distinguishes the quantum
world from the classical one, quantum entanglement has been employed in most
of the quantum information processing tasks (QIPTs) and has been paid close
attention to in wide field [1]. However, quantum entanglement can not cover
all the quantumness of correlations in a composite quantum system. It has
been shown that some QIPTs without any quantum entanglement might still
demonstrate quantum advantage if such QIPTs own quantum discord [2-9] which
was first introduced as the discrepancy between the generalizations of two
classically equivalent mutual information [10,11]. This could be one of the
potential reasons why quantum discord has attracted so many interests in the
past few years (see Ref. [12-27] and the references therein).

As an important branch of the researches on quantum correlation, the
quantification of quantum correlation is a hot topic [10,11,28-34]. The
information theoretic definition of quantum discord is found to be only
analytically calculated for some special states [10,11,15, 30,36]. The
geometric quantum discord can be calculated analytically for all $\left(
2\otimes d\right) $ -dimensional systems [28,29], but it confronts some
contradictions [37,38]. For example, 1) It could be increased by the
non-unitary evolution on the subsystem without measurements; 2) It could be
reduced by an extra product state. In order to avoid these unexpected
properties, some new measures based on the trace norm [33,39-40] are proposed  at the cost of losing the
computability including some
variational attempts [37,41,42] only covering 2).  It is fortunate that a recent progress [34] according to the
skew information [43] has effectively covered both the aspects, which shed
new light on the quantification of quantum correlation. However, much as the
quantification of entanglement is still restricted to the pure states and
low dimensional bipartite quantum states due to the complex optimization for
a general high dimensional quantum systems [1], so far it has still been an
open question how to provide a good quantification or even an effective
(fast, steadily, reliable, with machine precision) algorithm for the quantum
correlation in high dimensional systems.

In this Letter, we give a definition of quantum correlation pertaining to
arbitrary bipartite quantum systems based on the skew information. This
definition automatically inherits the good properties such as the
contractivity, so it is guaranteed to be a good measure. It is found to have
an interesting relation with quantum metrology . In addition, this
definition looks like that in Ref. [34], but it is quite different from it
even in pure states and the $\left (2\otimes d\right)$ -dimensional quantum
states. In particular, our definition for any dimensional state can be
converted to an existing and easy optimization question that can be fast,
steadily, reliably and effectively solved with machine precision by the well
developed technique. In this sense, we think our measure of quantum
correlation is even almost analytic. As a demonstration, we give the exact
expressions of quantum correlation for $\left
(2\otimes d\right)$%
-dimensional states, a type of positive partial transpose (PPT) states, the
high-dimensional Werner states and the Isotropic states, by which, on the
one hand, one will find the power of our definition of quantum correlation,
on the other hand, one will find the effectiveness of the proposed numerical
method.

\textit{The definition of quantum correlation}.-To begin with, we would like
to briefly introduce the skew information for a bipartite density matrix $%
\rho _{AB}$ and an observable $O$. It is defined by%
\begin{equation}
I\left( \rho _{AB},O\right) =-\frac{1}{2}Tr\left[ \sqrt{\rho _{AB}},O\right]
^{2}.
\end{equation}%
$I\left( \rho _{AB},O\right) $ has been employed in many fields [44-46], and
has many good properties [34]. For example, it vanishes if and only if $\rho 
$ and $O$ commute and it is positive in other cases; It doesn't increase
under classical mixing; In particular, if we select an observable $%
O=K_{A}\otimes \mathbf{1}_{B}$ with $K_{A}$ some observable on subsystem $A$%
, $I\left( \rho _{AB},O\right) $ is contractive under completely positive
and trace-preserving maps $\Phi $ on $B$, that is, $I\left( \left( \mathbf{1}%
_{A}\otimes \Phi \right) \rho _{AB},K_{A}\otimes \mathbf{1}_{B}\right) \leq
I\left( \rho _{AB},K_{A}\otimes \mathbf{1}_{B}\right) $.

In order to quantify the quantum correlation, Ref. [34] has required the
non-degenerate traceless observable $K_{A}$ (full rank) operated on
subsystem $A$. Here, we would like to restrict us to the \textit{rank-1}
local projectors. Let $\rho _{AB}$ be an $\left( m\otimes n\right) $%
-dimensional density matrix and suppose 
\begin{equation}
K_{k}=\left\vert k\right\rangle \left\langle k\right\vert \otimes \mathbf{1}%
_{n}
\end{equation}
with $\left\vert k\right\rangle $ in arbitrary orthonormal set $S$ of
subsystem $A$, we will be able to give our definition of quantum correlation
as follows.

\textbf{Definition. 1}.-The quantum correlation $\mathcal{Q}$ of $\rho _{AB}$
is defined by the minimal skew information induced by an group of
orthonormal projectors. This can be rephrased as 
\begin{equation}
\mathcal{Q}\left( \rho _{AB}\right) :=-\frac{1}{2}\min_{S}\sum_{k=0}^{m-1}Tr%
\left[ \sqrt{\rho _{AB}},K_{k}\right] ^{2},
\end{equation}%
with $K_{k}$ defined by Eq. (2).

\textbf{Proof. }In order to show $\mathcal{Q}\left( \rho _{AB}\right) $ is a
measure of quantum correlation, we have to prove that $\rho _{AB}$ is a
classical-quantum state as $\rho _{AB}=\sum \tilde{\lambda}_{k}\left\vert 
\tilde{k}\right\rangle \left\langle \tilde{k}\right\vert \otimes \varrho _{%
\tilde{k}}$ with $\left\vert \tilde{k}\right\rangle $ analogous to $%
\left\vert k\right\rangle $ in some set $\tilde{S}$ if and only if $\mathcal{%
Q}\left( \rho _{AB}\right) =0$.

Let $\rho _{AB}=\sum \tilde{\lambda}_{k}\left\vert \tilde{k}\right\rangle
\left\langle \tilde{k}\right\vert \otimes \varrho _{\tilde{k}}$, one can
always find such a $\tilde{k}$ that $\left[ \rho _{AB},K_{\tilde{k}}\right]
=0$ holds for all $\tilde{k}$. On the contrary, given an arbitrary $\rho
_{AB}$, if $\left[ \rho _{AB},K_{k_{1}}\right] =0$ for some particular $%
k_{1} $, we can write 
\begin{equation}
\rho _{AB}=\lambda _{k1}\left\vert k_{1}\right\rangle \left\langle
k_{1}\right\vert \otimes \varrho _{1}+\rho _{k_{1\bot }},
\end{equation}%
where $\rho _{k_{1\bot }}$ means that it can be completely expanded in the
orthogonal space of $\left\vert k_{1}\right\rangle \left\langle
k_{1}\right\vert $ and $\lambda _{k1}\geq 0$. If $\rho _{AB}$ given in Eq.
(4) continues to commuting with $K_{k_{2}}$ with $\left\langle k_{1}\right.
\left\vert k_{2}\right\rangle =0$, one can further write $\rho _{AB}$ as 
\begin{equation}
\rho _{AB}=\lambda _{k1}\left\vert k_{1}\right\rangle \left\langle
k_{1}\right\vert \otimes \varrho _{1}+\lambda _{k2}\left\vert
k_{2}\right\rangle \left\langle k_{2}\right\vert \otimes \varrho _{2}+\rho
_{k_{1\bot }\cap k_{2\bot }}.
\end{equation}%
If we $\left[ \rho _{AB},K_{k_{j}}\right] =0$ holds for all $k_{j}$ such
that $\sum_{j}\left\vert k_{j}\right\rangle \left\langle k_{j}\right\vert =%
\mathbf{1}_{m}$, one will draw the conclusion that 
\begin{equation}
\rho _{AB}=\sum_{j}\lambda _{k_{j}}\left\vert k_{j}\right\rangle
\left\langle k_{j}\right\vert \otimes \varrho _{j},
\end{equation}%
which is obviously a classical-quantum state. The proof is completed.\hfill$%
\blacksquare$

Next, one will easily find that the quantum correlation measure $\mathcal{Q}%
\left( \rho _{AB}\right) $ satisfies all the good properties that a measure
should meet.

(i) $\mathcal{Q}\left( \rho _{AB}\right) $ \textit{is invariant under local
unitary operations. }It is apparent that $I\left( \rho _{AB},O\right) $ is
invariant under local unitary operations, which is analogous to the proof in
Ref. [34].

(ii) $\mathcal{Q}\left( \rho _{AB}\right) $ \textit{is contractive under
competely positive and trace-preserving maps }$\Phi $\textit{\ on }$B$.%
\textit{\ }Since $I\left( \rho _{AB},K_{k}\right) $ given in Eq. (1) is
constractive for observable $K_{k}$, the same property is inherited by $%
\sum_{k}I\left( \rho _{AB},K_{k}\right) $. For an optimal set $\left\{
K_{k}\right\} $, $\mathcal{Q}\left( \rho _{AB}\right) \geq $ $%
\sum_{k}I\left( \left( \mathbf{1}_{m}\otimes \Phi \right) \rho
_{AB},K_{k}\right) \geqslant \mathcal{Q}\left( \left( \mathbf{1}_{m}\otimes
\Phi \right) \rho _{AB}\right) .$

(iii) $\mathcal{Q}\left( \rho _{AB}\right) $ \textit{is reduced to
entanglement for pure states. }Because $\mathcal{Q}\left( \rho _{AB}\right) $
is not changed by the local unitary operations, we can safely consider the
pure state in the form of Schmidt decomposition which is given by $%
\left\vert \chi \right\rangle _{AB}=\sum_{i=0}^{r-1}\mu _{i}\left\vert
ii\right\rangle _{AB}$ with $\mu _{i}$ the Schmidt coefficients and $r=\min
\{m,n\}$. Substitute $\left\vert \chi \right\rangle _{AB}$ into Eq. (3), one
will easily find that 
\begin{eqnarray}
&&\mathcal{Q}\left( \rho _{AB}\right) =1-\max_{S}\sum_{k=0}^{n-1}\left\vert
\sum_{i,j=0}^{r-1}\mu _{i}\mu _{j}\left\langle ii\right\vert \left(
\left\vert k\right\rangle \left\langle k\right\vert \otimes \mathbf{1}%
_{n}\right) \left\vert jj\right\rangle _{AB}\right\vert ^{2}  \notag \\
&&=1-\max_{S}\sum_{k=0}^{n-1}\left\vert \left\langle k\right\vert
\sum_{i=0}^{r-1}\mu _{i}^{2}\left\vert i\right\rangle _{A}\left\langle
i\right\vert \left. k\right\rangle \right\vert ^{2}  \notag \\
&&\geqslant 1-\sum_{k=0}^{r-1}\mu _{k}^{4}=1-Tr\varrho _{r}^{2},
\end{eqnarray}%
where $\varrho _{r}$ is the reduced density matrix of $\left\vert \chi
\right\rangle _{AB}$ and the "=" in Eq. (7) can always be satisfied if the
optimized set $S=\left\{ \left\vert i\right\rangle \right\} $.

\textit{The almost analytic expression for }$Q\left( \rho _{AB}\right) .$-
Based on the previous results, one can say that $\mathcal{Q}\left( \rho
_{AB}\right) $ is a good measure of quantum correlation. In the proceeding
part, we will convert the complex optimization question presented in Eq. (3)
into an existing and easy question, by which we will give the almost
analytic expression of $\mathcal{Q}\left( \rho _{AB}\right) .$

\textbf{Theorem 1.}-Let $\left\{ \left\vert i\right\rangle \right\} $ and $%
\left\{ \left\vert j\right\rangle \right\} $ denote two sets of orthonormal
bases of the subspace $B$ of the state $\rho _{AB}$, and let the Hermitian
matrices $A_{ij}=\left( \mathbf{1}_{m}\otimes \left\langle i\right\vert
\right) \sqrt{\rho _{AB}}\left( \mathbf{1}_{m}\otimes \left\vert
j\right\rangle \right) $, the quantum correlation $\mathcal{Q}\left( \rho
_{AB}\right) $ of $\rho _{AB}$ defined in Eq. (3) can be explicitly given by 
\begin{equation}
\mathcal{Q}\left( \rho _{AB}\right)
=1-\sum\limits_{i,j=0}^{n-1}\sum_{k=0}^{m-1}\left\vert \lambda
_{k}^{ij}\right\vert ^{2},
\end{equation}%
where $\lambda _{k}^{ij}$ means the $k$th joint eigenvalue of $A_{ij}$ which
is defined by $\left( U_{o}A_{ij}U_{o}^{\dag }\right) _{kk}$ with $U_{o}$
the joint diagonalizer of all the $A_{ij}$.

\textbf{Proof.} From Eq. (3), one will directly arrive at%
\begin{eqnarray}
&&\mathcal{Q}\left( \rho _{AB}\right) =\min_{S}\sum_{k=0}^{m-1}\left[ Tr\rho
_{AB}K_{k}^{2}-Tr\sqrt{\rho _{AB}}K_{k}\sqrt{\rho _{AB}}K_{k}\right]  \notag
\\
&&=1-\max_{S}\sum_{k=0}^{m-1}Tr\sqrt{\rho _{AB}}K_{k}\sqrt{\rho _{AB}}K_{k}.
\end{eqnarray}%
Substitute $K_{k}=\left\vert k\right\rangle \left\langle k\right\vert
\otimes \mathbf{1}_{n}$ and any orthonormal bases of subsystem $B$ into Eq.
(9), it follows that 
\begin{eqnarray}
\mathcal{Q}\left( \rho _{AB}\right)
&=&1-\max_{S}\sum\limits_{i,j=0}^{n-1}\sum_{k=0}^{m-1}Tr\sqrt{\rho _{AB}}%
\left( \left\vert k\right\rangle \left\langle k\right\vert \otimes
\left\vert i\right\rangle \left\langle i\right\vert \right)  \notag \\
&&\times \sqrt{\rho _{AB}}\left( \left\vert k\right\rangle \left\langle
k\right\vert \otimes \left\vert j\right\rangle \left\langle j\right\vert
\right)  \notag \\
&=&1-\max_{S}\sum\limits_{i,j=0}^{n-1}\sum_{k=0}^{m-1}\left\vert
\left\langle k\right\vert A_{ij}\left\vert k\right\rangle \right\vert ^{2} 
\notag \\
&=&1-\max_{U}\sum\limits_{i,j=0}^{n-1}\sum_{k=0}^{m-1}\left\vert
UA_{ij}U^{\dag }\right\vert _{kk}^{2},
\end{eqnarray}%
with 
\begin{equation}
A_{ij}=\left( \mathbf{1}_{m}\otimes \left\langle i\right\vert \right) \sqrt{%
\rho _{AB}}\left( \mathbf{1}_{m}\otimes \left\vert j\right\rangle \right) .
\end{equation}%
Thus, our calculation of the quantum correlation is directly changed into
the joint approximate diagonalization (JAD) of the series of matrices $%
A_{ij} $ [48,49]. Let $U_{o}$ be the joint diagonalizer of all the $A_{ij}$
such that the optimal value of Eq. (10) can be attained, and assume $\lambda
_{k}^{ij}=\left( U_{o}A_{ij}U_{o}^{\dag }\right) _{kk}$, one will easily
find that the final expression of $\mathcal{Q}\left( \rho _{AB}\right) $ can
be written as Eq. (8). It is obvious that if $[A_{ij},A_{kl}]=0$ holds for
all $A$, the question can be exactly solved. Of course, that these $A_{ij}$
commute with each other is just a sufficient condition for the exact
solution of Eq. (8), which will be seen from our latter examples. In
addition, the number of the matrices that need to be JAD can be reduced
further based on the Appendix. \hfill$\blacksquare$

Next we will briefly analyze why we say $\mathcal{Q}\left( \rho _{AB}\right) 
$ given in Eq. (8) is effective and almost analytic. At first, we would like
to claim that the effectivity of our result is completely attributed to the
well developed technique on the JAD (throughout this Letter, we especially
mean the Jacobi algorithm for JAD [48,49]), since we have succeeded in
converting the original particular optimization into such an existing JAD
question. In particular, we emphasize that by these well developed
techniques, especially the Jacobi algorithm which is also used to
diagonalize a single matrix, the JAD can be solved as steadily, reliably,
fast and perfectly as the diagonalization of a single matrix [49,50]. From
the point of practical applications of view, we would like to say that our
result in Eq. (8) is almost analytic, which could be intuitive if $\lambda
_{k}^{ij}$ were the exact eigenvalues of some particular matrix. In the
researches on the quantum correlation including quantum entanglement and
quantum discord, it is usual to accept that the result is analytic, if it
can be given by some eigenvalues of any given matrices. However, in
practical operations, these eigenvalues are usually calculated (especially
for large matrices) by computers with some default precision (or machine
precision). The technique of JAD is completely the generalization of that
for a single matrix. For example, the Jacobi algorithm has the complete same
principle as that for a single matrix [48,50]. So the JAD is completely on
the same level as the diagonalization of a single matrix, so is their
precision, efficiency, and so on. Quantitively, the JAD of Eq. (8) needs at
most $\frac{m(m-1)n^2}{2}$ Givens rotations for one ergodicity of the
entries of the matrix in Jacobi algorithm [49], while the diagonalization of 
$\rho_{AB}$ needs $\frac{mn(mn-1)}{2}$ rotations for one ergodicity [50].
Thus from the practical applications, we can think our result is almost
analytic in the sense of the "analytic diagonalization of a single matrix $%
\rho_{AB}$".

\textit{Relation with quantum metrology.-} Before proceeding, we will show
that our quantum correlation in Definition 1 connects some quantum metrology
scheme in an interesting way. Let $\left\{ \left\vert k\right\rangle
\right\} $ denote the group of optimal orthonormal set of projectors that
achieves the exact value of $\mathcal{Q}\left( \rho _{AB}\right) $. Assume
the state $\rho _{AB}$ is a probing state with subsystem $A$ undergoing a
unitary transformation which endows some unknown phases $\varphi _{k}$ on $%
\rho _{AB}$ by $\rho _{\vec{\varphi}}=e^{-iH(\vec{\varphi})}\rho _{AB}e^{iH(%
\vec{\varphi})}$ with $H(\vec{\varphi})=\sum_{k}\varphi _{k}\left\vert
k\right\rangle \left\langle k\right\vert \otimes \mathbf{1}_{n}$. We aim to
estimate these $\varphi _{k}$ one by one by $N$ runs of detection with high
precision quantified by the uncertainty of the estimated phase $\varphi
_{k}^{est}$ $\left( \delta \varphi _{k}\right) ^{2}=\left\langle \frac{%
\varphi _{k}^{est}}{\partial \left\langle \varphi _{k}^{est}\right\rangle
/\partial \varphi _{k}}-\varphi _{k}\right\rangle $ [51,52]. This variance $%
\delta \varphi _{k}$, for an unbiased estimator, is bounded by the quantum
Cramer-Rao bound $\left( \delta \varphi _{k}\right) ^{2}\geq \frac{1}{NF_{Qk}%
}$ that can be attained asymptotically by the projective measurements in the
basis of the symmetric logarithmic derivative operator and the maximum
likelihood estimation, where $F_{Qk}$ is the quantum Fisher information
subject to the phase $\varphi _{k}$ [51-54]. What we would like to emphasize
is that $F_{Qk}=-Tr\left[ \sqrt{\rho _{AB}},K_{k}\right] ^{2}$, so one can
easily find that $\sum_{k}\frac{1}{N\left( \delta \varphi _{k}\right) ^{2}}%
\leq \sum_{k}F_{Qk}=2\mathcal{Q}\left( \rho _{AB}\right) $. That is, our
quantum correlation measure characterizes the contributions of all the
inverse variances of the estimated phases.

\textit{The applications.-}From the following, one will find that $\mathcal{Q%
}\left( \rho _{AB}\right) $ for some states can be analytically solved,
whilst these examples will demonstrate the effectiveness of the JAD method
in the calculation for high-dimensional systems and illustrate the perfect
consistency between the strictly analytic solutions and the almost analytic
ones obtained by the JAD method.

\begin{figure}[tbp]
\centering
\includegraphics[width=0.6\columnwidth]{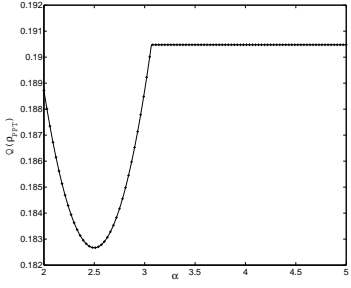}
\caption{The quantum correlation $\mathcal{Q}\left(\protect\rho_{PPT}\right)$
for the PPT states versus $\protect\alpha$. The solid line and the "+" line
correspond to the numerical expression of Theorem 1 and strictly analytical
expression of Eq. (8). The transition from the separable state to bound
state is exactly at $\protect\alpha=3$, but the sudden change point of
quantum correlation is at about $\protect\alpha=3.066885$. When $\protect%
\alpha$ goes beyond the point, the quantum correlation will keep invariant.
This figure shows that our theorem 1 is effective and almost analytic in
contrast to the strict solution.}
\end{figure}

\textit{(a) Qubit-qudit states.-}As a comparison with the previous jobs, we
will first consider the quantum correlation\textit{\ }of a $\left( 2\otimes
d\right) $-dimensional state. For such a state $\rho _{AB}$, Eq. (3) can be
rewritten as 
\begin{eqnarray}
\mathcal{Q}\left( \rho _{AB}\right)  &=&-\frac{1}{2}\min_{S}\sum_{k=0}^{1}Tr%
\left[ \sqrt{\rho _{AB}},K_{k}\right] ^{2}  \notag \\
&=&-\min_{S}Tr\left[ \sqrt{\rho _{AB}},K_{0}\right] ^{2}.
\end{eqnarray}%
Since any pure state can be expanded in the Bloch representation, one can
always write $K_{0}$ as%
\begin{equation}
K_{0}=\frac{1}{2}\left( \mathbf{1}_{2}+\vec{n}\cdot \vec{\sigma}\right)
\otimes \mathbf{1}_{d}
\end{equation}%
with $\sum n_{i}^{2}=1$. Substitute Eq. (13) into Eq. (12), one will arrive
at 
\begin{eqnarray}
\mathcal{Q}\left( \rho _{AB}\right)  &=&\frac{1}{2}-\frac{1}{2}\max_{\vec{n}%
}\sum_{ij}Trn_{i}T_{ij}n_{j}  \notag \\
&=&\frac{1}{2}\left( 1-\upsilon _{\max }\right), 
\end{eqnarray}%
where $\upsilon _{\max }$ is the maximal eigenvalue of the matrix $T$ with 
\begin{equation}
T_{ij}=Tr\sqrt{\rho _{AB}}\left( \sigma _{i}\otimes \mathbf{1}_{n}\right) 
\sqrt{\rho _{AB}}\left( \sigma _{j}\otimes \mathbf{1}_{n}\right) .
\end{equation}
Eq. (14) happened to be the half of that in Ref. [34].

\textit{(b) $\left( 3\otimes 3\right) $ -dimensional PPT states.}-Let's
consider such a PPT state given by [55] 
\begin{equation}
\rho _{PPT}=\frac{2}{7}\left\vert \Phi \right\rangle _{3}\left\langle \Phi
\right\vert +\frac{\alpha }{7}\rho _{+}+\frac{5-\alpha }{7}\rho _{-},\alpha
\in \lbrack 2,4],
\end{equation}%
where $\left\vert \Phi \right\rangle _{m}=\frac{1}{\sqrt{m}}%
\sum_{k=0}^{m-1}\left\vert kk\right\rangle $ and $\rho _{+}=\frac{1}{3}%
\sum\limits_{k=0}^{2}\left\vert k,k\oplus 1\right\rangle \left\langle
k,k\oplus 1\right\vert $ and $\rho _{-}=\frac{1}{3}\sum\limits_{k=0}^{2}%
\left\vert k\oplus 1,k\right\rangle \left\langle k\oplus 1,k\right\vert $
with "$\oplus "$ the modulo-3 addition. Note that, only when $\alpha \in
(3,4]$, $\rho _{PPT}$ is entangled. If $\alpha \leq 3$, $\rho _{PPT}$ is
separable. But if $4<\alpha \leq 5$, $\rho _{PPT}$ is not a PPT state, but a
free entangled state. For integrity, we also consider this type free
entangled states here. It is interesting that, $\mathcal{Q}\left( \rho
_{PPT}\right) $ can be analytically solved for $\alpha \in \lbrack 2,5]$,
which is given by%
\begin{equation}
\mathcal{Q}\left( \rho _{PPT}\right) =\left\{ 
\begin{array}{cc}
\frac{21-\sqrt{6\left( 5-\alpha \right) }-\sqrt{6\alpha }-3\sqrt{\alpha
\left( 5-\alpha \right) }}{31.5}, & 2\leq \alpha \leq N_{T} \\ 
\frac{4}{21}, & N_{T}<\alpha \leq 5%
\end{array}%
\right. ,
\end{equation}%
with $N_{T}=\frac{15+\sqrt{136\sqrt{94}-1307}}{6}=3.066885$. The numerical
results based on our theorem 1 is plotted in Fig. 1, which shows the perfect
consistency between our theorem 1 and the strict analytic expression. In
particular, we can analytically find the sudden change point of quantum
correlation near the critical point of the separable state and the bound
entangled state.
\begin{figure}[tbp]
\centering
\includegraphics[width=1\columnwidth]{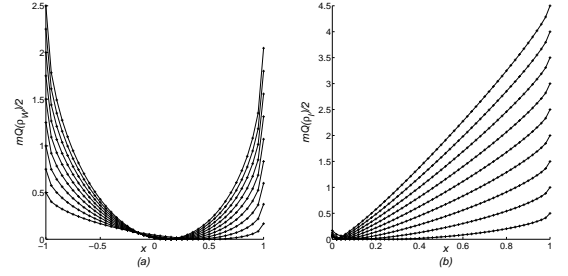}
\caption{The quantum correlation $\mathcal{Q}\left( \protect\rho _{W}\right) 
$ and $\mathcal{Q}\left( \protect\rho _{I}\right) $ for the Werner states
and Isotropic states versus $x$. The solid line corresponds to the numerical
expression given by Theorem 1 and the '+' marks the line produced by the
strictly analytical solutions given by Eq. (8). The lines from the bottom
(see the right side) to the top correspond to $m=2,3,\cdot \cdot \cdot ,10$.
All show the perfect consistency. }
\end{figure}

\textit{(c) Werner states and Isotropic states in }$\left( m\otimes m\right) 
$ dimension\textit{.}-Besides the above examples, our quantum correlation
measure $\mathcal{Q}\left( \cdot \right) $ for both the $\left( m\otimes
m\right) $ -dimensional Isotropic states and Werner states [53] can be
analytically calculated. Thus they can serve as important examples that show
the effectivity of $\mathcal{Q}\left( \cdot \right) $ for larger systems.
The Werner state can be written as%
\begin{equation}
\rho _{W}=\frac{m-x}{m^{3}-m}\mathbf{1}_{m^{2}}+\frac{mx-1}{m^{3}-m}V,x\in
\lbrack -1,1],
\end{equation}%
with $V=\sum_{kl}\left\vert kl\right\rangle \left\langle lk\right\vert $ the
swap operator. This state has no quantum correlation if and only if $x=\frac{%
1}{m}$. Through a simple algebra, one can have the analytic expression of
the quantum correlation as follows. 
\begin{equation}
\mathcal{Q}\left( \rho _{W}\right) =\frac{m-x-\sqrt{m^{2}-1}\sqrt{1-x^{2}}}{%
2(1+m)}.
\end{equation}%
From Eq. (19), one will also find that $\mathcal{Q}\left( \rho _{W}\right)
=0 $ for $x=\frac{1}{m}$. Analogously, we also plot $\mathcal{Q}\left( \rho
_{W}\right) $ based on Eq. (19) and Eq. (8), respectively, in Fig. 2 (a)
which shows the perfect consistency. The isotropic state can be given by%
\begin{equation}
\rho _{I}=\frac{1-x}{m^{2}-1}\mathbf{1}_{m^{2}}+\frac{m^{2}x-1}{m^{2}-1}%
\left\vert \Phi \right\rangle \left\langle \Phi \right\vert ,x\in \lbrack
0,1],
\end{equation}%
with $\left\vert \Phi \right\rangle =\frac{1}{\sqrt{m}}\sum_{k=0}^{m-1}\left%
\vert kk\right\rangle $. Based on our definition, we can analytically obtain%
\begin{equation}
\mathcal{Q}\left( \rho _{I}\right) =\frac{1-2\sqrt{m^{2}-1}\sqrt{x(1-x)}%
+\left( m^{2}-2\right) x}{m(1+m)}.
\end{equation}%
It is obvious that $x=\frac{1}{^{m^{2}}}$ will lead to $\mathcal{Q}\left(
\rho _{I}\right) =0$, which is consistent to Ref. [15]. As a comparison, we
plot $\mathcal{Q}\left( \rho _{I}\right) $ given by Eq. (8) and Eq. (21),
respectively, in Fig. 2 (b) which shows the perfect consistency again.

\textit{Conclusions and Discussion.}-We have presented a new definition of
quantum correlation for any bipartite quantum system with some good
properties. In particular, this definition can lead to an effective and even
almost analytic expression for any states. As applications, we have found
that the quantum correlations of many quantum states can be strictly
analytically solved, which also provides a direct support for the
effectivity of our theorem.

Finally, we would like to emphasize that the JAD technique plays an
important role in our job. Whether it can induce other contributions to the
relevant researches such as quantum correlation of other forms, quantum
entanglement measure etc. is worthy of our forthcoming efforts.

\textit{Acknowledgements}-This work was supported by the National Natural
Science Foundation of China, under Grants No.11375036 and No. 11175033, and by the Fundamental
Research Funds of the Central Universities, under Grant No. DUT12LK42.

\begin{equation}
\sum\limits_{i,j=0}^{n-1}\sum_{k=0}^{m-1}\left\vert UA_{ij}U^{\dag
}\right\vert _{kk}^{2}=\sum_{k=0}^{m-1}\left[ \left( U\otimes U\right)
S\left( U^{\dag }\otimes U^{\dag }\right) \right] _{kk}^{kk},
\end{equation}%
where $S=\sum_{i,j=0}^{n-1}A_{ij}\otimes A_{ij}^{\dag }$. Based on the
Kronecker Approximation technique [56], one can always express $S$ as 
\begin{equation}
S=\sum_{k=0}^{r-1}B_{k}\otimes B_{k}^{\dag },
\end{equation}%
where $r$ is the rank of the matrix $\tilde{M}=V_{s}(SV_{s})^{T_{2}},$ the
superscript $T_{2}$ denotes partial transposition on the second space [57], $%
V_{s}$ is the swap operator defined by $V_{s}\left\vert \varphi
\right\rangle \left\vert \psi \right\rangle =\left\vert \psi \right\rangle
\left\vert \varphi \right\rangle $\ for any two states $\left\vert \psi
\right\rangle $ and $\left\vert \varphi \right\rangle $ with the same
dimension. If the singular value decomposition of $\tilde{M}$ is given by 
\begin{equation}
\tilde{M}=U\Sigma V^{\dagger }=\sum_{i=0}^{r-1}\sigma _{i}u_{i}v_{i}^{\dag },
\end{equation}%
where $u_{i}$ is the $i$th columns of the unitary matrix $U$; $\Sigma $ is a
diagonal matrix with elements $\sigma _{i}$ in the decreasing order, one
will easily find that $Vec(B_{k})=\sqrt{\sigma _{k}}u_{k}$. Thus the $%
\mathcal{Q}\left( \rho _{AB}\right) $ can be rewritten as%
\begin{equation}
\mathcal{Q}\left( \rho _{AB}\right)
=1-\max_{U}\sum_{j=0}^{r-1}\sum_{k=0}^{m-1}\left\vert UB_{j}U^{\dag
}\right\vert _{kk}^{2}
\end{equation}%
with $\lambda _{k}^{t}$ the $k$th joint eigenvalue of $B_{t}$.

\end{document}